\documentclass[12pt]{iopart}

\usepackage[english,activeacute]{babel}
\usepackage{graphicx}
\usepackage{amsfonts}
\usepackage{amssymb}
\usepackage{epsf}
\usepackage{latexsym}
\usepackage{epsfig,multicol}
\include{titlesec-conf}
\include{titletoc-conf} 

\newcommand{\unit}{\leavevmode\hbox{\small1\kern-3.6pt\normalsize1}}

\def\neumass{m_{\tilde\chi_1^0}}
\newcommand{\crosssec}{\sigma_{\tilde\chi^0_1-p}}

\def\neut{\tilde\chi_1^0}

\def\lsim{\raise0.3ex\hbox{$\;<$\kern-0.75em\raise-1.1ex\hbox{$\sim\;$}}}
\def\gsim{\raise0.3ex\hbox{$\;>$\kern-0.75em\raise-1.1ex\hbox{$\sim\;$}}}
\def\nmh{{\sc nmhdecay}}

\newcommand{\captions}{\sf\caption}

\begin{document}
\hspace*{10.1cm}FTUAM 07/05\\
\hspace*{11cm}IFT-UAM/CSIC-07-11

\title {Neutralino Dark Matter in the NMSSM}

\author{Daniel E. L\'opez-Fogliani} 

\address{Departamento de F\'isica Te\'orica C-XI \& Instituto de F\'isica Te\'orica C-XVI, Universidad Aut\'onoma de Madrid, Cantoblanco, E-28049 Madrid, Spain.}
\ead{lopezd@delta.ft.uam.es}
\begin{abstract}
We analyze the direct detection of neutralino dark matter in the framework of the Next-to-Minimal Supersymmetric Standard Model. Taking into account all the available constraints from LEPII and Tevatron, including bounds from the muon anomalous magnetic moment and flavor constraint as $b \to s \gamma $, we compute the neutralino-nucleon cross section and compare the results with the sensitivity of dark matter detectors. We also study the relic abundance of neutralinos, comparing it with WMAP observations. We show that a light singlet-like Higgs can escape accelerator constraints allowing for scenarios with large cross sections, and a very interesting phenomenology.
\end{abstract}


\section{Introduction}
Weakly Interacting Massive Particles (WIMP's) are excellent candidates for dark matter, and can in principle be detected via elastic interaction with  a nucleus. Actually, a large number of experiments are devoted to the direct detection of WIMP's.
Collaborations like DAMA/LIBRA, EDELWEISS, CDMS, XENON 1 Tonne of Ge/Xe, etc. are running and/or in progress \cite{review}.

Supersymmetric (SUSY) theories with R-parity conservation offer an appealling candidate for dark matter: the neutralino as the lightest supersymmetric particle (LSP).
The theoretical predictions for the direct detection of neutralino dark matter in the framework of the Next-to-Minimal Supersymmetric Standard Model (NMSSM) has been analyzed recently  \cite{Cerdeno:2004xw,emideo}, 
and we will use these analyses in what follows.
Let us remark that the NMSSM provides a solution to the so called $\mu$-problem, and makes less severe the "Higgs-little fine tuning" problem in the MSSM, via the introduction of a singlet field. 
Although the symmetries of the NMSSM may give rise to the possibility of a cosmological domain wall problem \cite{Abel1},
 this can be avoided by the introduction of suitable non-renormalisable operators 
\cite{Abel2}.
On the other hand, the singlet field mixes with the MSSM Higgses, and the singlino field mixes with the MSSM neutralinos, thus offering a richer phenomenology. In particular, it is possible to have very light neutralinos. In addition, it is also possible to have a very light Higgs which is experimentally viable for a sufficiently high singlet composition.
The NMSSM superpotential is
$W=
\epsilon_{ij} \left(
Y_u \, H_2^j\, Q^i \, u +
Y_d \, H_1^i\, Q^j \, d +
Y_e \, H_1^i\, L^j \, e \right)
- \epsilon_{ij} \lambda \,S \,H_1^i H_2^j +\frac{1}{3} \kappa S^3\,$
and, when the scalar component of $S$ acquires a VEV, an effective interaction $\mu H_1 H_2$ is generated, with $\mu \equiv \lambda \langle S \rangle$.
We refer to \cite{Cerdeno:2004xw} for a detailled analysis of the Higgs scalar potential and minimization conditions. 


%
\section{Constraints on the parameter space
}
In addition to ensuring the presence of a minimum of the potential, many other constraints, both theoretical and experimental, must be imposed on the parameter space of the NMSSM: $\lambda, \, \kappa,\, \tan \beta,\, \mu$, the trilinear soft parameters  $ A_\lambda, \, A_\kappa, \, A_{U,D,E}$, the gaugino masses $M_{1,2,3}$, and the scalar masses $m_{Q,L,U,D,E}$. Let us remark that in our computation all these parameters are specified at low scale.
%
 A comprehensive analysis of the low-energy NMSSM phenomenology can be obtained using the \nmh~2.0 code~\cite{Ellwanger:2005dv}. 
We also impose flavor constraints, being the most relevant $b \to s \gamma $.
The recent experimental World average for the branching ratio (BR) reported by the Heavy Flavor Averaging Group is~\cite{bsg_exp} ${\rm BR}^{\rm exp}(b\to s \gamma)=(3.55 \pm 0.27)\times10^{-4}$ .
We include in our code the calculation of the $b\to s\gamma$  branching ratio at the NLO order, following the results 
of~\cite{kagan-neubert}.
%
Finally, in our analysis we will also take into account the constraints coming
from the SUSY contributions to the muon anomalous magnetic moment, $a_{\mu}=(g_{\mu}-2)$.
Note that the current average experimental value is $ a^{\rm exp}_{\mu}=11\, 659\, 208(6)\times 10^{-10}\, $~\cite{g-2_exp}.


\section{Results and discussion}
\begin{figure}[t]
 \hspace*{.6cm}
 \includegraphics[scale=.5,angle=0]{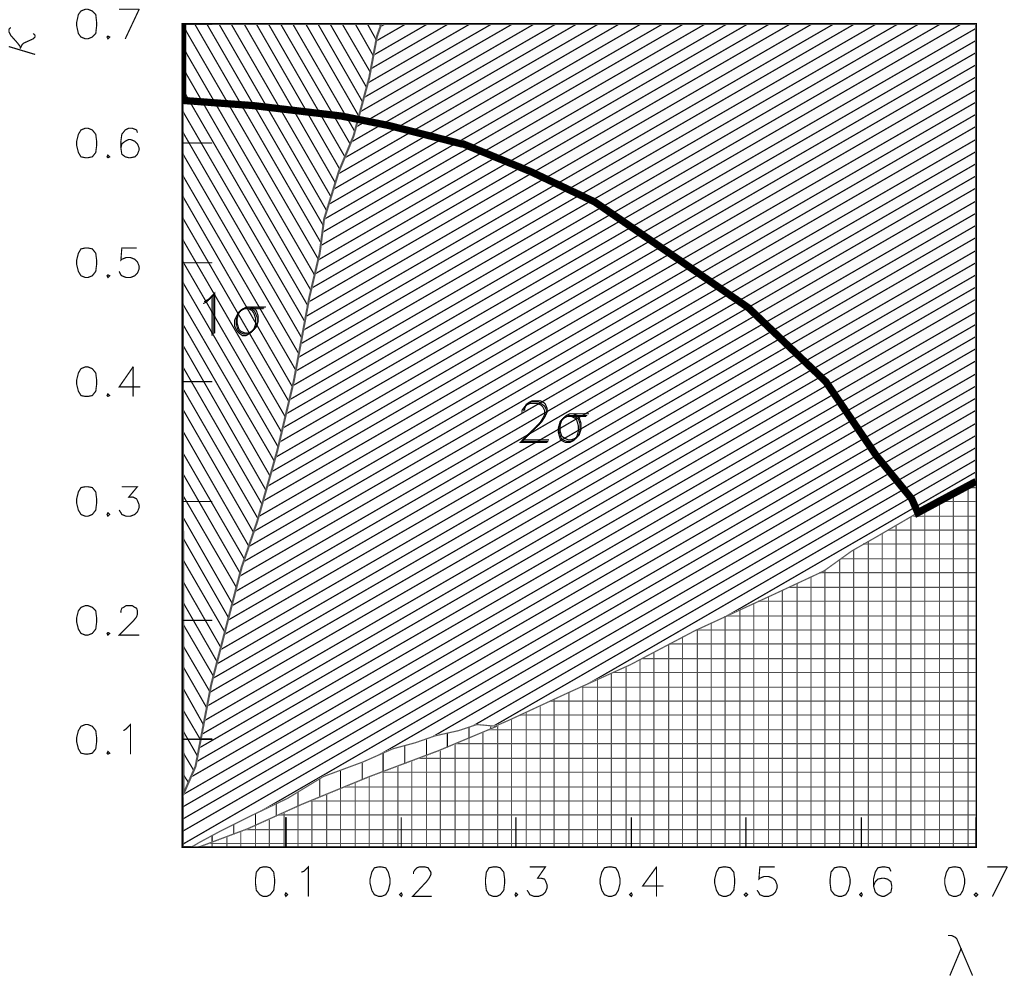}
\hspace*{.3cm}
  \includegraphics[scale=.5,angle=0]{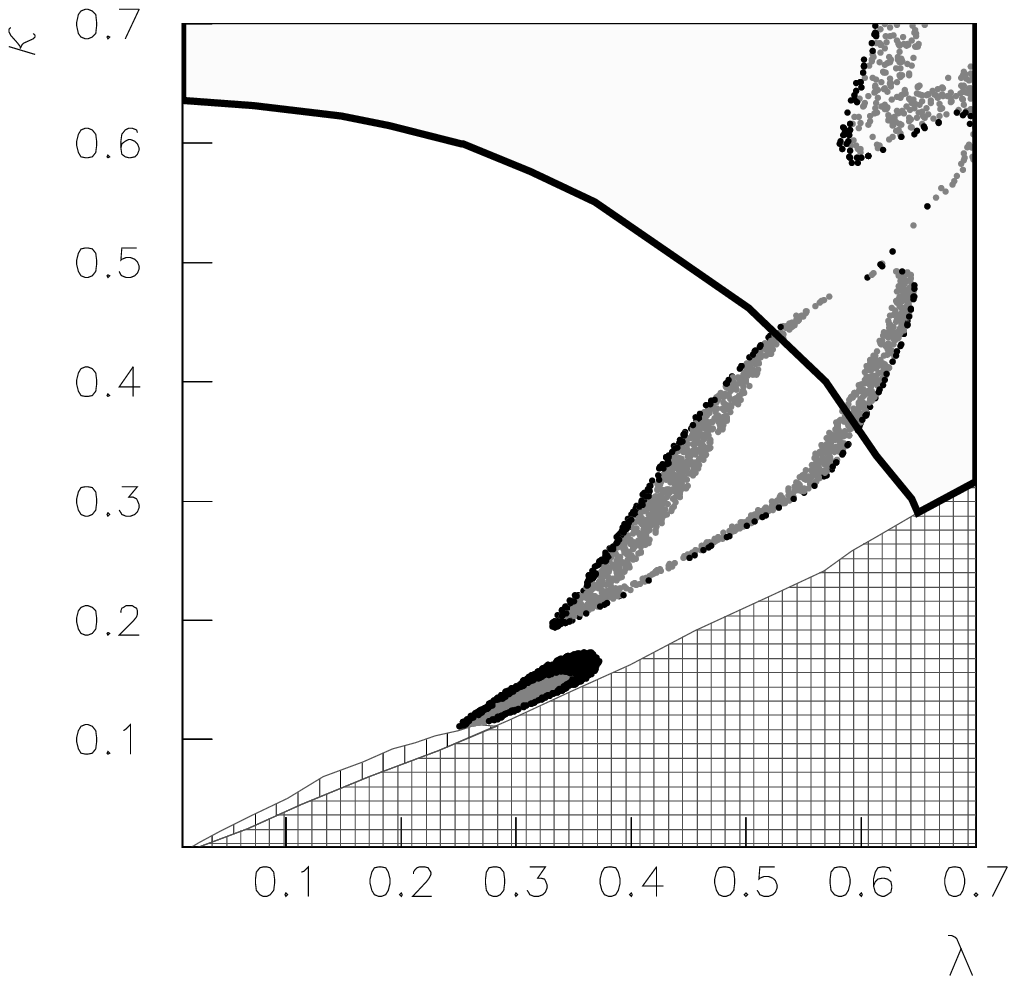}
\captions{$(\lambda,\kappa)$ parameter space for $\tan\beta=5 $, $A_\lambda=400 $~GeV, $A_\kappa=-200$~GeV, $\mu=130$~GeV.  In both cases the gridded area is excluded by tachyons and the region above the thick black line is excluded by the occurrence of a Landau Pole below the GUT scale (on the right this region appears in light gray). On the left we plot the  $1\sigma$ and $2\sigma$ regions for the BR($b \to s \gamma$). On the right we plot the neutralino relic density where black points are in agreement with 
the recents  observations mentioned in the text, $0.095 < \Omega h^2 < 0.112$ ~\cite{wmap}. For comparison points in agreement with old observations, $0.1 \lesssim \Omega h^2 \lesssim 0.3$ \cite{review} are shown in gray.
The vertically ruled area corresponds to the occurrence of unphysical minima.}
\label{fig1}
\end{figure}

The relevant parameters at low scale are $\lambda$, $\kappa$, $\mu$, $\tan \beta$, $A_\lambda $, $A_\kappa $  the soft gaugino masses $M_1$, $M_2$, $M_3$. In addition we assume the relation $M_1=\frac{M_2}{2}=\frac{M_3}{6}$ that mimics the GUT unification. 
We are interested in the region of the parameter space that enables cross section in the sensitivity range of detectors. We further want to be within the 2$\sigma$ deviations of the central values of the anomalous magnetic moment of the muon. For this reason we choose the relevant slepton parameters $M_E=150$~GeV, $M_L=150$~GeV and the trilinear $A_E=-2500$~GeV.
In the quark sector we take $M_{Q,D,U}=A_{D}=1$~TeV, $A_U=2.4$~TeV. We consider that all flavor mixing comes from the CKM matrix. In this situation the dominant channel for the $b \to s \gamma$ is mediated by the charged Higgs. The BR($b \to s\,\gamma$) closely follows the behavior of the charged Higgs mass, which in the NMSSM is given at tree level~\footnote[1]{We note that in our computation of the Higgs masses radiative corrections are taken into account~\cite{Ellwanger:2005dv}.} by
$m_{H^\pm}^2\;=\;\frac{2 \,\mu^2}{\sin(2 \beta)}\,
 \frac{\kappa}{\lambda}\,-\, v^2
 \,\lambda^2 \,+\, \frac{2 \,\mu \,A_\lambda}{\sin(2 \beta)}
 \,+\, M_{W}^2\,$.
From the above, we expect that smaller values of BR($b \to s\,\gamma$) should be obtained for large $\kappa/\lambda$ (for positive values of $\kappa$) or small $\kappa/\lambda$ (for $\kappa < 0$). In general, smaller values of the BR($b \to s \gamma$) will be also associated to larger values of the product $\mu \,A_\lambda$. On the other hand, we have also included in the computation the constraints on the relic abundance coming from the combination of the recent WMAP3 observations  \cite{wmap} with CMB, large scale structure, and Hubble constant measurements, $ 0.095 < \Omega h^{2} < 0.112 $ .
With respect to the relic abundance, one of the regions where $\Omega h^2$ is potentially within the WMAP range is roughly associated with the band where $M_1 \approx \mu$ \cite{Belanger:2005kh}. 
 To fullfil the above constraints we also choose  $M_1=160$~GeV, $A_\lambda=400$~GeV, $\mu=130$~GeV, $A_\kappa=-200$~GeV, $\tan(\beta)=5$, as a first example. In Fig.\ref{fig1} we present the BR($b \to s \gamma$) (on the left) and the relic abundance (on the right), both in the $(\lambda,\kappa)$ plane.

\begin{figure}[!h]
\hspace*{.6cm}
  \includegraphics[scale=.5,angle=0]{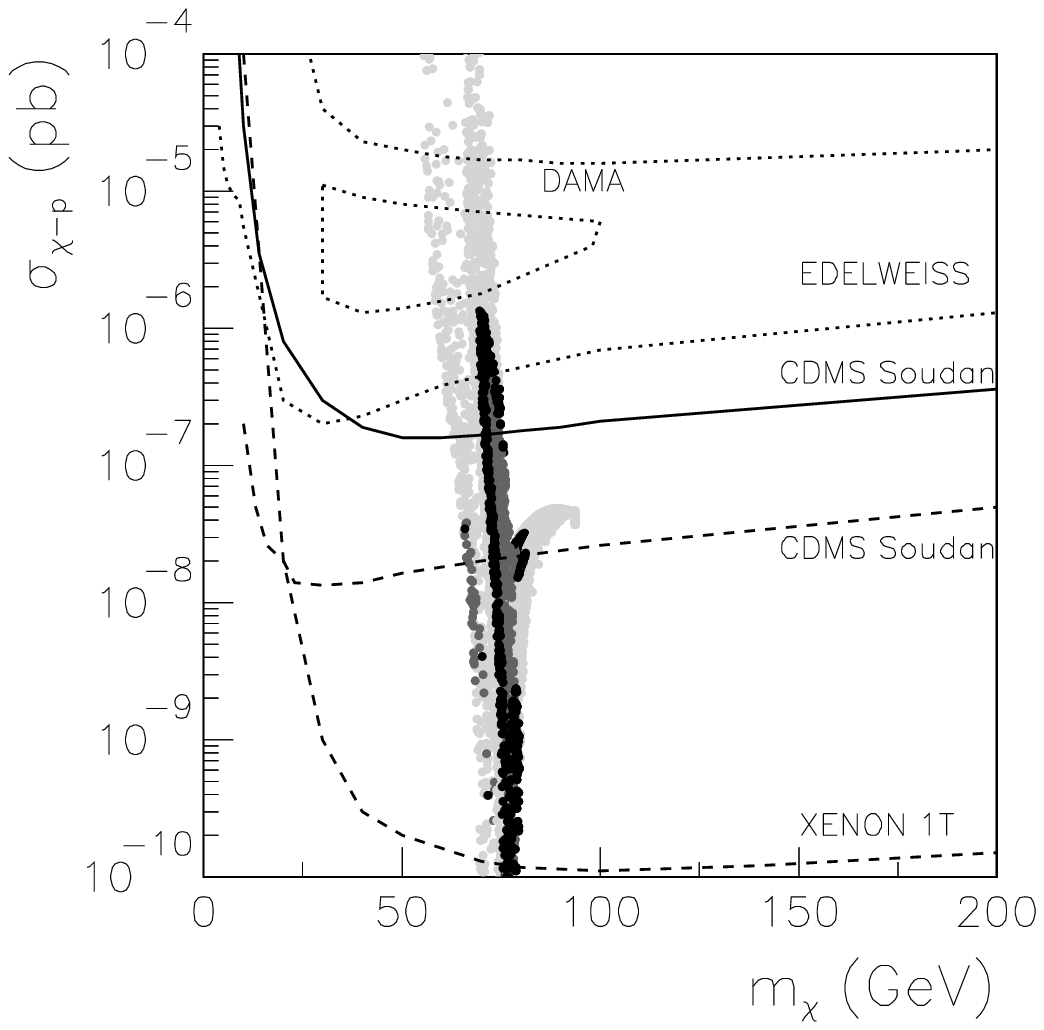}
 \hspace*{.3cm}
  \includegraphics[scale=.5,angle=0]{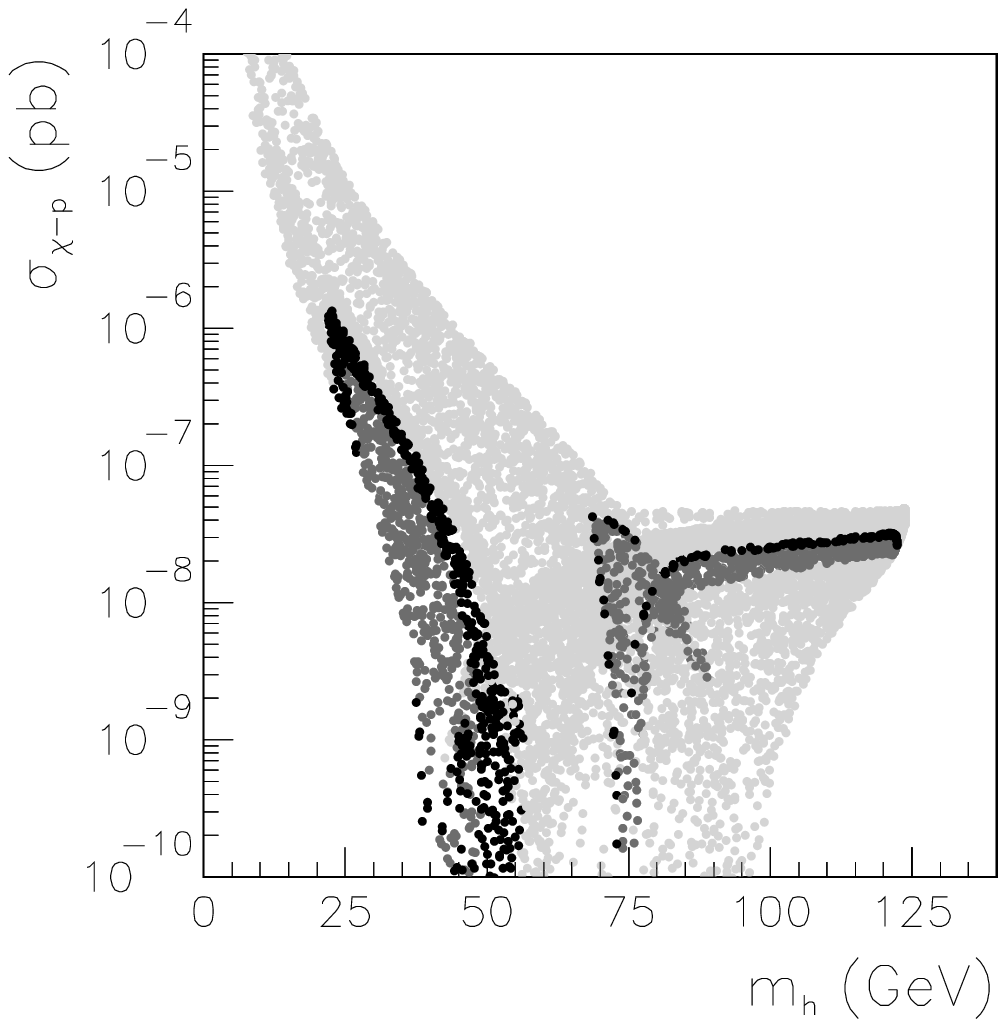}
  \captions{Scatter plot of the scalar neutralino-nucleon cross section, $\crosssec$, as a function of  the neutralino mass, $\neumass$ (on the left), and as a function of  the lightest scalar Higgs mass, $m_{h_1^0}$ (on the right). We take $A_\lambda=400$~GeV, $\mu=130$~GeV, $A_\kappa=-200$~GeV, and $\tan\beta=5$. Black (gray) dots correspond to points fulfilling all the experimental constraints including the relic density in the range $0.095 < \Omega h^2 < 0.112$ ($0.1 \lesssim \Omega h^2 \lesssim 0.3$). Light grey dots represent those excluded. On the left the sensitivities of present and projected experiments are also depicted with solid and dashed lines, respectively. The large (small) area bounded by dotted lines is allowed by the DAMA  experiment when astrophysical uncertainties are (are not) taken into account.
  }
  \label{fig2}
\end{figure}

%
%
We find that for large values of $\kappa$ with respect to $\lambda$, the LSP is a mixture of Bino and Higgsinos.
This, together with a comparatively large mass, ensures an easy annihilation into gauge boson pairs, so
that the relic density is negligible. For intermediate values of $\kappa$, we observe that both the LSP and the lightest Higgs become more singlino and singlet respectively. In this example the singlino composition of the neutralino is always less than 35 \%. As the neutralino mass decreases, some annihilation channels become
kinematically forbidden, such as annihilation into a pair of $Z$ or $W$ bosons when $\neumass<M_Z$ or $\neumass<M_W$, respectively. 
In these cases, $\Omega h^2$ can be large (above the WMAP3 range).
Moreover, the spectrum is such that one can encounter $m_{h^0_2} \approx 2 m_{\tilde \chi^0_1}$
resonances. 
%
Finally, for the regions with very small values of $\kappa$ the Higgs becomes lighter than the neutralino and new annihilation channels (the most important being $\neut\neut\to h_1^0h_1^0$ and $Zh_1^0$) are available for the neutralino, thus decreasing its relic density.
On Fig \ref{fig2} we plot on the left the neutralino-nucleon cross section as a function of the neutralino, and on the right as a function of the lightest Higgs mass. Large cross section can be obtained for small Higgs masses $m_{h^0} \approx  50$~GeV.
%
%
\begin{figure}[!h]
 \hspace*{.6cm}
  \includegraphics[scale=.5,angle=0]{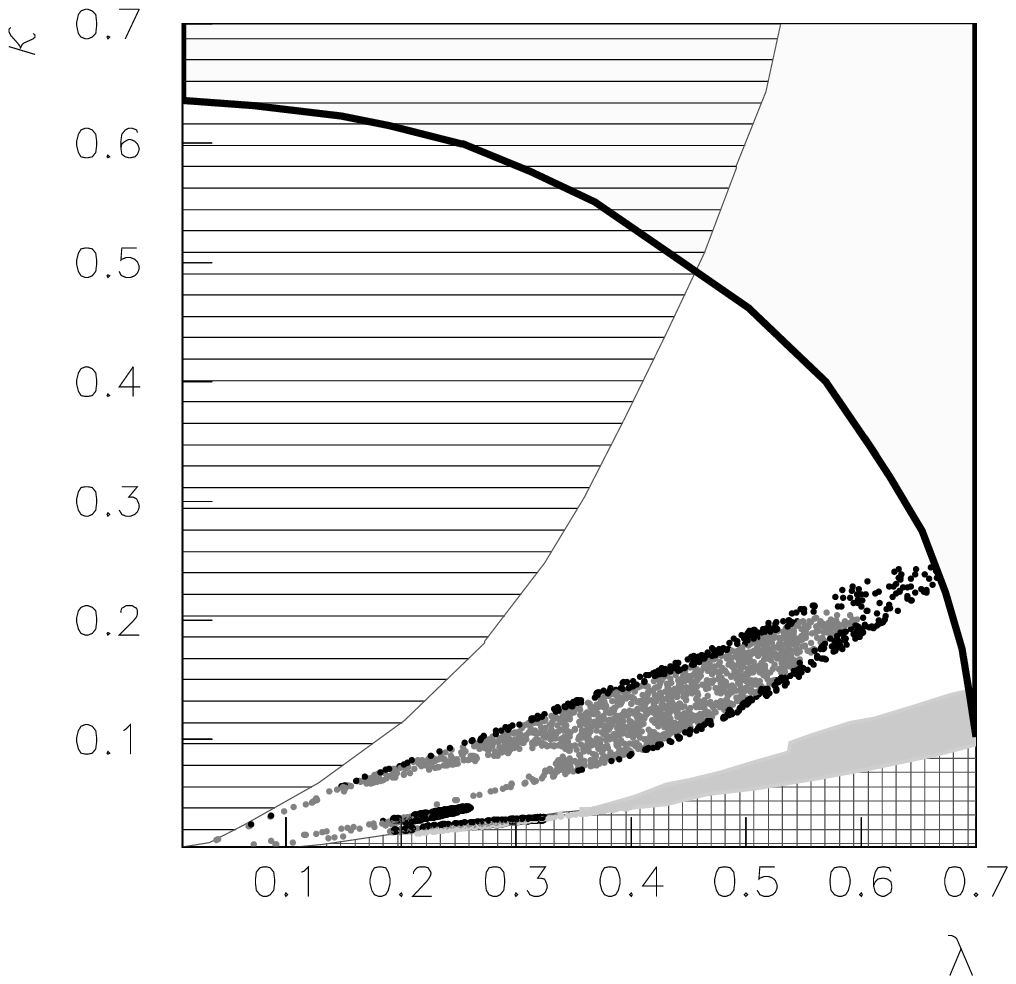}
  \hspace*{.3cm}
\includegraphics[scale=.5,angle=0]{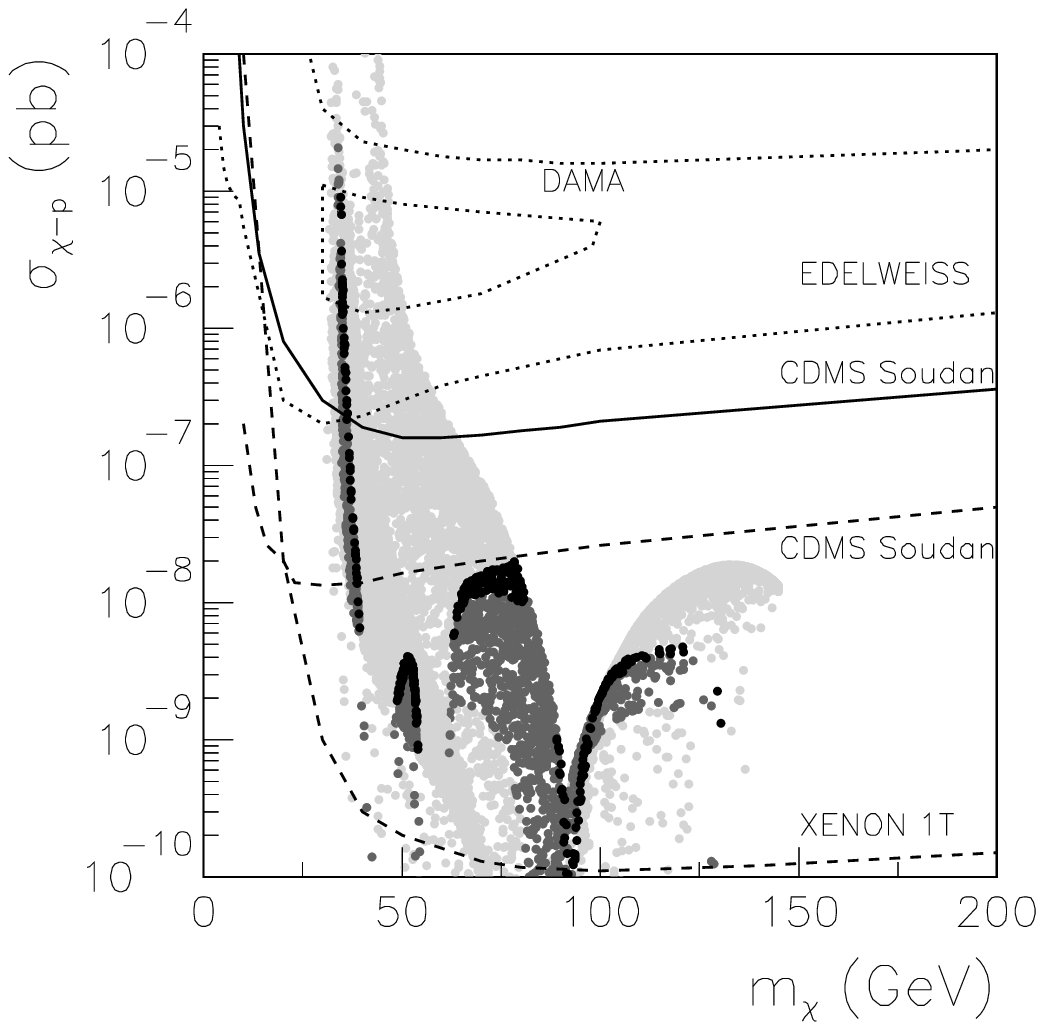}
  \captions{For $M_1=330$~GeV, $A_\lambda=570$~GeV, $\mu=160$ GeV, $A_\kappa=-60$ GeV, and $\tan\beta=5$. On the left the  same conventions as on the right of Fig. 1. In addition horizontal lines represent excluded points where the LSP is not the neutralino. On the right the same as on the left of Fig. 2.
  }\label{fig3}
\end{figure}
On Fig. \ref{fig3} we show another case in which it is possible to have a light neutralino with a large component of singlino. 
We display the results for the relic density and the cross section.
In this case all the $(\lambda, \kappa)$ plane is within the $2 \sigma $ region of accuracy for BR($b \to s \gamma$). On the other hand $a_\mu $ is slightly beyond the $2 \sigma $ region. On the left of Fig. \ref{fig3} we observed a region with horizontal lines corresponding to $m_{\tilde \chi^0_1} $ were the neutralino is not the LSP.

\section{Conclusions}
Working in the NMSSM, we have  computed the theoretical predictions for the scalar neutralino-proton cross section $\crosssec$, and compared it with the sensitivities of present and projected dark matter experiments.  In the computation we have taken into account all available experimental constraint from LEP and Tevatron on the parameter space, constraints coming from B and K physics, and the muon anomalous magnetic moment. We have also checked the correspondence between the theoretical and observational values of the relic density. We have found that large values of $\crosssec$, even within the reach of present dark matter detectors, can be obtained in regions of the parameter space. This is essentially due to the exchange of very light Higgses.

\noindent {\bf Acknowledgments}

L-F gratefully acknowledges C. Mu\~noz, A. M. Teixeira, D. G. Cerde\~no and E. Gabrielli for their valuable help and collaboration. L-F also wishes to thank the organizers of IRGAC2006.
This work was supported by the MEC under contract FPA2006-01105; by the Comunidad de Madrid under proyecto HEPHACOS S-0505/ESP-0346; 
and by the EU under the ENTApP Network of the ILIAS Project No. RII3-CT-2004-506222.


\section*{References}
\begin{thebibliography}{10}
\bibitem{review} For a recent review see, C. Mu\~noz, {\it Int.
J. Mod. Phys.} {\bf A19} (2004) 3093-3170.



%
%
%

\bibitem{Cerdeno:2004xw}
D.~G.~Cerde\~no, C.~Hugonie, D.~E.~L\'opez-Fogliani, C.~Mu\~noz and A.~M.~Teixeira, JHEP {\bf 0412} (2004) 048.
\bibitem{emideo} D.~G.~Cerde\~no, E.~Gabrielli, D.~E.~L\'opez-Fogliani, C.~Mu\~noz and A.~M.~Teixeira, arXiv:hep-ph/0701271.
\bibitem{Abel1} S.~A.~Abel, S.~Sarkar and P.~L.~White,
{\it Nucl.\ Phys.\ } {\bf B454} (1995) 663.
\bibitem{Abel2} S.~A.~Abel, 
{\it Nucl.\ Phys.\ } {\bf B480} (1996) 55.
%
%
\bibitem{Ellwanger:2005dv}
U.~Ellwanger and C.~Hugonie, Comput. Phys. Commun. {\bf 175} (2006) 290-303.
\bibitem{bsg_exp}
The Heavy Flavour Averaging Group, http://www.slac.stanford.edu/xorg/hfag/;
S. Eidelman, Phys. Lett. {\bf B 592} (2004) 1.
\bibitem{kagan-neubert}
A. L. Kagan and M. Neubert, Eur. Phys. J. {\bf C7} (1999) 5.
\bibitem{g-2_exp}
G.W. Bennett {\it et al.} [Muon g-2 Collaboration], Phys. Rev. Lett.
{\bf 92} (2004) 161802.
\bibitem{wmap} D. N. Spergel et al.,
arXiv:astro-ph/0603449.
\bibitem{Belanger:2005kh}
G.~Belanger, F.~Boudjema, C.~Hugonie, A.~Pukhov and A.~Semenov, JCAP {\bf 0509} (2005) 001.

\end{thebibliography}
\end{document}